# A High-Resolution Transmission Line Model with De-embedding Structure for Ultralow Contact Resistivity Extraction

Xuanyu Jia, Hongxu Liao, Ming Li, School of Integrated Circuits, Peking University

**Abstract—** In this article, we present a contact resistivity extraction method calibrated using a de-embedding structure, called High-Resolution Transmission Line Model (HR-TLM). HR-TLM has the similar infrastructure with Refined TLM (RTLM) or Refined-Ladder TLM(R-LTLM), but is optimized for calibration methods. Its advantage lies in maintaining low $\rho_c$ extraction accuracy while significantly reducing the impact of structural process errors. According to the error analysis model, we verify that the extraction accuracy of HR-TLM based on R-LTLM can reach $10^{-9}$ $\Omega\cdot cm^2$ at micron scale lithography precision.

## I. Introduction

As CMOS dimensions continue to scaling down, the contact resistance $R_c$ in the source/drain (S/D) region has become a major limiting factor on the current drivability, surpassing the effects caused by channel resistance $R_{ch}$ [1]-[3]. For the sub-7 nm technology node, ultralow contact resistivities $\rho_c$ (below $1\times10^{-9}$ $\Omega\cdot cm^2$) are eagerly required for high-performance devices [4]. Therefore, efforts have focused on accurate ultralow $\rho_c$ extraction models.

Nowadays, various low-$\rho_c$ extraction methods are used in academia. Refined Transmission Line Model (RTLM) and Nano Transmission Line Model (nano-TLM) emerge as an enhancement over the conventional TLM by reducing non-idealities like current crowding effects and the resistive contributions of metal layers, which are especially suitable for nanoscale dimension contacts [5]. Multi-Ring Circular Transmission Line Model (MR-CTLM), as an advancement of the Circular Transmission Line Model (CTLM), requires only one photolithography step for a simple fabrication process while maintaining an extraction accuracy of $10^{-9}$ $\Omega\cdot cm^2$ [2], [4]. Ladder Transmission Line Model (LTLM) effectively eliminates the influence of parasitic metal resistances from contact metals and access electrodes by employing a ladder-like resistor network [6],[7].

The above extraction models enable ultra-low contact resistivity measurements, but they need very high structural dimensional accuracy, typically requiring nanoscale lithographic precision. Small process errors in the test structure can significantly bias the extraction results. Based on this issue, X. Sun et al. proposed a Refined Ladder Transmission Line Model(R-LTLM), relieving the impact of process variation during the formation of contact area patterns and contact spacings [8]. However, the practical implementation shows the stability of the extraction results of R-LTLM is still insufficient at micrometer lithography precision, and can be further optimized.

In this work, we propose a novel High-Resolution Transmission Line Model(HR-TLM), which is an optimization scheme based on RTLM or R-LTLM, using a zero-space de-embedding structure for calibration. The advantage of HR-TLM is that it eliminates the parasitic resistance of the metal layer and access electrodes to ensure the extraction accuracy, and reduces the requirement of process accuracy. HR-TLM can achieve a $\rho_c$ extraction accuracy below $10^{-9}$ $\Omega\cdot cm^2$ at micron scale lithography precision.

## II. Description of HR-TLM based on RTLM

The test structure of RTLM is shown in **Fig. 1(a)**, where the current is injected between Pad A and D and the voltage between Pad B and C is measured. The total resistance $R_T$ is

$$R_t(L_s) = 2R_A + 2R_C + R_M = 2 \times \frac{R_{shm} \times L_c}{W} + 2R_C + \frac{R_{shs} \times L_s}{W} \tag{1}$$

Where $R_{shm}$ and $R_{shs}$ are the sheet resistance of metal and substrate layers, $L_s$ is the length of substrate layer, $L_c$ is the effective length of metal layer, W is the test section width, $R_C$ is the contact resistance, which can be expressed as

$$R_C \approx \frac{R_{shs} \times L_t}{W} = \frac{R_{shs}}{W}\sqrt{\frac{\rho_c}{R_{shs} + R_{shm}}} \tag{2}$$

Where $L_t = \sqrt{\frac{\rho_c}{(R_{shs}+R_{shm})}}$ is the transfer length. The meaning of Equation (2) is to equate the contact resistance to the resistance over the transfer length.

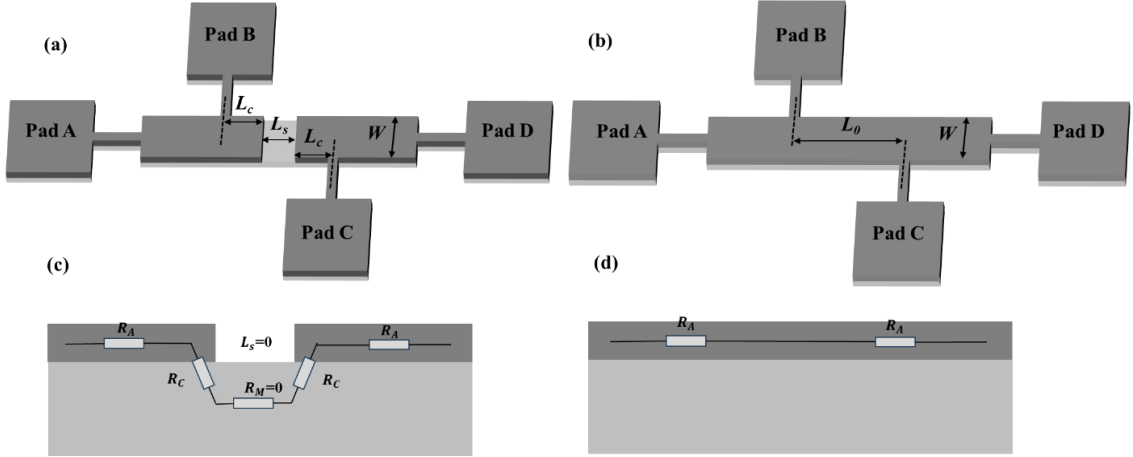

**Fig. 1.** (a) test structure of RTLM and HR-TLM. (b)the de-embedding structure for calibration. (c) theoretical resistance structure when $L_s = 0$. (d) the resistance of the de-embedded structure.

HR-TLM has similar basic structure as RTLM, but features an additional de-embedding structure for calibration, as shown in **Fig. 1(b)**. In HR-TLM, we keep $L_0 = L_s + 2L_c$ as a constant, i.e., $L_c$ varies with $L_s$. The resistance of the de-embedded structure is shown in **Fig. 1(d)**, which can be approximated as current flowing entirely through the metal layer. In Equation (1), when $L_s = 0$, the corresponding theoretical resistance structure is shown in **Fig. 1(c)**. It differs from **Fig. 1(d)** in the presence of a partition in the metal layer, which in turn introduces additional contact resistance. Then introduce $R_{HR-TLM}(L_s)$,

$$R_{HR-TLM}(L_s) = R_t(L_s) - R_{no\ spacer} = \frac{(R_{shs} - R_{shm}) \times L_s}{W} + 2R_C \tag{3}$$

Where $R_{no\ spacer}$ is the measured resistance of the de-embedded structure. When $L_s = 0$,

$$R_{HR-TLM}(0) = 2R_C = \frac{2R_{shs}}{W}\sqrt{\frac{\rho_c}{R_{shs} + R_{shm}}} \tag{4}$$

With the slope $S = \frac{(R_{shs}-R_{shm})}{W}$ and intercept $R_{HR-TLM}(0)$ obtained from the linear fit, we can calculate $\rho_c$.

$$\rho_c = (\frac{R_{HR-TLM}(0) \cdot (R_{shs} - R_{shm})}{2S})^2 \times \frac{R_{shs} + R_{shm}}{R_{shs}^2} \tag{5}$$

Some approximate calculations are utilized in the methods presented in this section, so the extraction accuracy needs to be further verified experimentally.

## III. Description of HR-TLM based on R-LTLM

In the conventional refined-ladder transmission line model (R-LTLM)[8], the total resistance can be described as

$$R_t(L_g) = S \cdot L_g + 2R_{c,LTLM} + \frac{R_{shs}L_0}{W} + 2(R_A + R_C), \text{ as shown in \textbf{Fig. 2(c)}} \quad (6)$$

$$S = -\frac{R_{shs}^2}{W(R_{shs} + R_{shm})} \quad (7)$$

$$R_{c,LTLM} = \frac{R_{shs}^2}{W(R_{shs} + R_{shm})} \cdot L_t \quad (8)$$

Where $L_g$ is the length of the ladder region (**Fig. 2(a)**), $L_0 = 2L_s + L_g$ is kept as constant, i.e., $L_s$ varies with $L_g$. However, the practical implementation shows that variations in the dimensions of the total gap ($L_0$) may introduce large systematic errors in contact resistivity extraction when using the no-ladder structure (corresponding ladder length $L_g$=0) as the de-embedding part. To improve the extraction accuracy and numerical stability, we propose the novel HR-TLM to eliminate both the parasitic metal resistance and the systematic errors from $L_0$ variation.

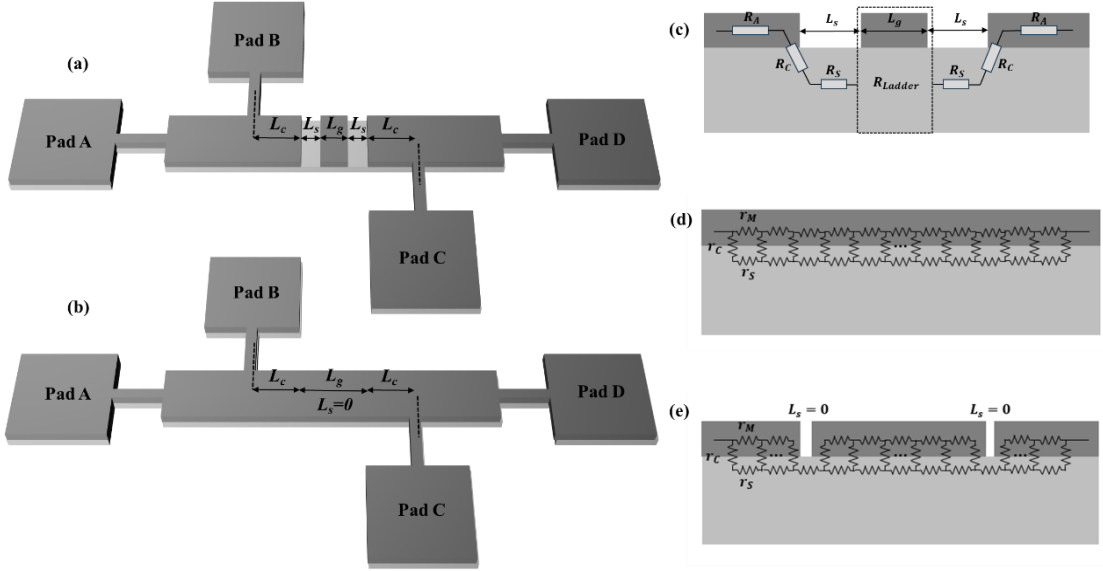

Fig. 2. (a) test structure of R-LTLM and HR-TLM. (b)the de-embedding structure for calibration. (c) the resistance distribution schematic of the test structure. (d) the differential resistance network of the de-embedded structure. (e) the differential resistance network of the theoretical structure when $L_g = 0$.

The basic structure is also the same as that of R-LTLM, while introducing an additional de-embedded structure with full metal layer coverage, as shown in **Fig. 2(b)**. Its resistance is shown in **Fig. 2(d)**, with the corresponding theoretical resistance for $L_g = 0$ shown in **Fig. 2(e)**. Unlike section II, the difference in the resistances here cannot be simply considered as $4R_C$, but needs to be solved by the differential resistance network.

For the network in **Fig. 2(d)**, we can obtain from Kirchhoff's loop law that

$$\frac{I_m(x)R_{shm}}{W} - \frac{I_s(x)R_{shs}}{W} = -\frac{\partial V(x)}{\partial x} \quad (9)$$

$$\frac{\partial I_m(x)}{\partial x} + \frac{V(x)W}{\rho_c} = 0 \quad (10)$$

$$I_m(x) + I_s(x) = I_0 \quad (11)$$

Where $I_m(x)$ and $I_s(x)$ are the currents in the metal and the semiconductor along the x-direction,

respectively. $I_0$ is the total current. $V(x)$ is the potential difference between the upper and lower layers at position x. The differential resistances in **Fig. 2** are denoted as $r_M = \frac{R_{shm}}{W} dx$, $r_S = \frac{R_{shs}}{W} dx$, $r_C = \frac{\rho_c}{W dx}$. With Equation (9)-(11), we can get the differential equation for $I_m(x)$ and its generalized solution,

$$\frac{\partial^2 I_m(x)}{\partial x^2} = \frac{(R_{shs} + R_{shm})I_m(x) - R_{shs}I_0}{\rho_c} \tag{12}$$

$$I_m(x) = A\sinh\left(\frac{x}{L_t}\right) + B\cosh\left(\frac{x}{L_t}\right) + \frac{R_{shs}I_0}{R_{shs} + R_{shm}} \tag{13}$$

The boundary conditions are

$$I_m(0) = I_0 \tag{14}$$

$$I_m\left(\frac{L_0}{2}\right) \approx I_m(\infty) = \frac{R_{shs}I_0}{R_{shs} + R_{shm}} \tag{15}$$

Equation (15) holds provided that $\frac{L_0}{2} \gg L_t$ while $I_m(x)$ always converges, and the structure is left-right symmetric [8]. Then we have

$$I_m(x) = \frac{R_s I_0}{R_s + R_m} + \frac{R_m I_0}{R_s + R_m} e^{-\frac{x}{L_t}}, \quad \left(0 < x < \frac{L_0}{2}\right) \tag{16}$$

The total resistance is calculated from Equation (16), which is equal to the total potential drop across the metal layer divided by the total current $I_0$:

$$R_{no\ spacer} = \frac{2}{I_0}\int_0^{\frac{L_0}{2}} \frac{I_m(x)R_{shm}}{W} dx = \frac{R_{shs}R_{shm}(2L_c + L_0)}{W(R_{shs} + R_{shm})} + \frac{2R_{shm}^2 L_t}{W(R_{shs} + R_{shm})} \tag{17}$$

For the network in **Fig. 2(e)**, we can divide it into three parts by the two partitions. The middle part is similar to Equation (17), and only $R_{shs}$ and $R_{shm}$ need to be interchanged. The left and right parts are symmetrical. For the left part of the differential resistance network, we can write the same Kirchhoff equation as Equation (9)-(11), as well as the corresponding differential equation and the general solution:

$$\frac{\partial^2 I_s(x)}{\partial x^2} = \frac{(R_{shs} + R_{shm})I_s(x) - R_{shm}I_0}{\rho_c} \tag{18}$$

$$I_s(x) = C\sinh\left(\frac{x}{L_t}\right) + D\cosh\left(\frac{x}{L_t}\right) + \frac{R_{shm}I_0}{R_{shs} + R_{shm}} \tag{19}$$

The new boundary conditions are

$$I_s(0) = 0, \quad I_s(L_c) = I_0 \tag{20}$$

This boundary condition holds approximately, because $R_{shs} \gg R_{shm}$, and the current flows into the substrate almost entirely at $x = L_c$. Then we have

$$I_s(x) = \frac{R_{shm}I_0}{R_{shs} + R_{shm}} - \frac{R_{shm}I_0}{R_{shs} + R_{shm}} e^{-\frac{x}{L_t}} + \frac{2R_{shs}I_0}{R_{shs} + R_{shm}} e^{-L_c/L_t} \sinh\left(\frac{x}{L_t}\right) \tag{21}$$

This part of the resistance corresponds to $(R_A + R_C)$ in Equation (6).

$$R_A + R_C = \int_0^{L_c} \frac{I_s(x)R_{shs}}{WI_0} dx \approx \frac{R_{shs}R_{shm}L_c}{W(R_{shs} + R_{shm})} + \frac{(R_{shs}^2 - R_{shs}R_{shm})L_t}{W(R_{shs} + R_{shm})} \tag{22}$$

The total resistance of the network in **Fig. 2(e)** is

$$R_T(L) = \frac{2R_{shs}R_{shm}L_c}{W(R_{shs}+R_{shm})} + \frac{R_{shs}R_{shm}L_0}{W(R_{shs}+R_{shm})} + \frac{(4R_{shs}^2 - 2R_{shs}R_{shm})L_t}{W(R_{shs}+R_{shm})} \quad (23)$$

Equation (6) can be written more precisely as

$$R_t(L_g) = S \cdot L_g + \frac{R_{shs}L_0}{W} + \frac{R_{shs}R_{shm}L_c}{W(R_{shs}+R_{shm})} + \frac{(4R_{shs}^2 - 2R_{shs}R_{shm})L_t}{W(R_{shs}+R_{shm})} \quad (24)$$

With Equation (17) and (24), we have

$$R_{HR-TLM}(L_g) = R_t(L_g) - R_{no\ spacer}$$
$$= S \cdot (L_g - L_0) + \frac{(4R_{shs}^2 - 2R_{shs}R_{shm} - 2R_{shm}^2)L_t}{W(R_{shs}+R_{shm})} \quad (25)$$

When $L_g = L_0$,

$$R_{HR-TLM}(L_0) = \frac{(4R_{shs}^2 - 2R_{shm}^2 - 2R_{shs}R_{shm})L_t}{W(R_{shs}+R_{shm})} \quad (26)$$

With the slope $S$ and intercept $R_{HR-TLM}(L_0)$ obtained from the linear fit, we can calculate $\rho_c$.

$$\rho_c = \left(\frac{R_{HR-TLM}(L_0)}{2S}\right)^2 (R_{shs}+R_{shm}) \left(\frac{R_{shs}^2}{2R_{shs}^2 - R_{shm}^2 - R_{shs}R_{shm}}\right)^2 \quad (27)$$

Compared to R-LTLM, the test result of HR-TLM is more stable and credible at micron scale lithography accuracy, while maintaining the extraction accuracy. HR-TLM is optimized for process error sensitivity mainly in two ways. The first is the variation in the total space of the metal layers ($L_0$), which can have a huge impact on subsequent calculations as a baseline value. For R-LTLM, if $L_0$ varies by $\delta L_0$, the total resistance will change as

$$\frac{\delta R_T(L_g=0)}{\delta L_0} = \frac{R_{shs}}{W} \quad (28)$$

For HR-TLM, we have

$$\frac{\delta R_{no\ spacer}}{\delta L_0} = \frac{R_{shs}R_{shm}}{W(R_{shs}+R_{shm})} \quad (29)$$

$$\frac{\delta R_{no\ spacer}}{\delta R_T(L_g=0)} = \frac{R_{shm}}{R_{shs}+R_{shm}} \ll 1 \quad (30)$$

Equation (30) shows that HR-TLM structure significantly reduces the effect of $L_0$ variations.

The second optimization is that the de-embedded structure in R-LTLM ($L_g = 0$) is mainly from the substrate resistance of $L_s$ part. Since $R_{shs} \gg R_{shm}$, small deviations in the thickness and shape of the substrate can cause a large drift to the total calibration resistance, which in turn affects the subsequent extraction. The new de-embedding structure which is mainly from metal layer resistance can optimize the error well.

Differentiating Equation (27), we can obtain the error analysis model of HR-TLM.

$$\frac{\Delta\rho_c}{\rho_c} = sqrt[\left(2\frac{\Delta R_{HR-TLM}(L_0)}{R_{HR-TLM}(L_0)}\right)^2 - \left(2\frac{\Delta S}{S}\right)^2$$
$$+ \left(\left(\frac{R_{shs}}{R_{shs}+R_{shm}} - \frac{2(R_{shs}R_{shm}+2R_{shm}^2)}{2R_{shs}^2 - R_{shs}R_{shm} - R_{shm}^2}\right) \times \frac{\Delta R_{shs}}{R_{shs}}\right)^2$$
$$+ \left(\left(\frac{R_{shm}}{R_{shs}+R_{shm}} + \frac{2(R_{shs}R_{shm}+2R_{shm}^2)}{2R_{shs}^2 - R_{shs}R_{shm} - R_{shm}^2}\right) \times \frac{\Delta R_{shm}}{R_{shm}}\right)^2] \quad (31)$$

Simplifying with the approximation condition $R_{shs} \gg R_{shm}$, we have

$$\Delta\rho_c = \left[\frac{R_{HR-TLM}(L_0)}{2S}\right]^2 \times sqrt[\left(\frac{(R_{shs}+R_{shm})}{2R_{HR-TLM}(L_0)}\Delta R_{HR-TLM}(L_0)\right)^2$$

$$+ \left(\frac{(R_{shs} + R_{shm})}{2S}\Delta S\right)^2 + \left(\frac{\Delta R_{shs}}{4}\right)^2 + \left(\frac{\Delta R_{shm}}{2}\right)^2 ] \quad (32)$$

In our practical test at micron scale lithography accuracy, we have $R_{shs}$ and $R_{shm}$ in orders of 100 Ω/□ and 10 Ω/□, with $\Delta R_{shs}$ and $\Delta R_{shm}$ at $10^{-1}$ Ω/□. In linear fitting of $R_{HR-TLM}(L_g)$, the coefficient of determination $r^2$ can reach 0.9999, with $\Delta R_{HR-TLM}(L_0)$ at 0.5 Ω, and $\Delta S/S$ lower than 1%. According to the above data and Equation (32), we have $d\rho_c$ in the order of $10^{-10}$ Ω/cm². The credible value of $\rho_c$ extracted from our experiment can reach 1E-9 Ω/cm². If the process conditions can be further improved, the extraction accuracy of HR-TLM may be even better.

**IV. Conclusion**

An HR-TLM method is proposed and developed for the extraction of $\rho_c$ in the sub-$10^{-9}$ Ω/cm² regime. The test structure is based on RTLM and R-LTLM, introducing a zero-spacer de-embedding structure as optimization. HR-TLM can achieve credible contact resistivity extraction of 1E-9 Ω/cm² at micron scale lithography conditions, significantly reducing the structure's dependence on process accuracy.